\documentclass[final,authoryear, 3p,times,twocolumn]{elsarticle}
\usepackage[colorlinks=true,linkcolor=black, citecolor=blue, urlcolor=blue]{hyperref}

\usepackage{textcomp}
\usepackage{comment}
\usepackage{lineno}
\usepackage{amsmath}
\usepackage{orcidlink}
\usepackage{subcaption}
\newcommand{\scarlet}{{\sc scarlet}}
\newcommand{\tractor}{{\sc The Tractor}}
\newcommand{\astrophot}{{\sc Astrophot}}
\newcommand{\scarlettwo}{{\sc scarlet\oldstylenums{2}}}

\usepackage{amssymb}
\usepackage{lipsum}

\journal{Astronomy $\&$ Computing (accepted)}

\begin{document}

\begin{frontmatter}



\title{Disentangling transients and their host galaxies with \scarlettwo{}: \\ A framework to forward model multi-epoch imaging} 

 
\author[inst1]{Charlotte Ward$^{\orcidlink{0000-0002-4557-6682}}$}
\ead{charlotte.ward@princeton.edu}
\author[inst1,inst2]{Peter Melchior$^{\orcidlink{0000-0002-8873-5065}}$}
\author[inst1]{Matt L. Sampson$^{\orcidlink{0000-0001-5748-5393}}$}
\author[inst5]{Colin J. Burke$^{\orcidlink{0000-0001-9947-6911}}$}
\author[inst1]{Jared Siegel$^{\orcidlink{0000-0002-9337-0902}}$}
\author[inst1]{Benjamin Remy$^{\orcidlink{0000-0002-0978-5612}}$}
\author[inst1]{Sufia Birmingham$^{\orcidlink{0009-0003-7044-9751}}$}
\author[inst4]{Emily Ramey$^{\orcidlink{0000-0003-0447-8426}}$}
\author[inst3]{Sjoert van Velzen$^{\orcidlink{0000-0002-3859-8074}}$}

\affiliation[inst1]{organization={Department of Astrophysical Sciences, Princeton University},
            city={Princeton},
            postcode={08544}, 
            state={NJ},
            country={USA}}

\affiliation[inst2]{organization={Center for Statistics and Machine Learning, Princeton University},
            city={Princeton},
            postcode={08544}, 
            state={NJ},
            country={USA}}
\affiliation[inst5]{organization={Department of Astronomy, Yale University},
            addressline={266 Whitney Avenue}, 
            city={New Haven},
            postcode={06511}, 
            state={CT},
            country={USA}}
\affiliation[inst4]{organization={Department of Astronomy, University of California, Berkeley},
            addressline={}, 
            city={Berkeley},
            postcode={94720}, 
            state={CA},
            country={USA}}
\affiliation[inst3]{organization={Leiden Observatory, Leiden University},
            addressline={Postbus 9513, 2300 RA}, 
            city={Leiden},
            country={the Netherlands}}
            
\begin{abstract}
Many science cases for wide-field time-domain surveys rely on accurate identification and characterization of the galaxies hosting transient and variable objects. In the era of the Legacy Survey of Space and Time (LSST) at the Vera C. Rubin Observatory the number of known transient and variable sources will grow by orders of magnitude, and many of these sources will be blended with their host galaxies and neighboring galaxies. A diverse range of applications -- including the classification of nuclear and non-nuclear sources, identification of potential host galaxies in deep fields, extraction of host galaxy spectral energy distributions without requiring a transient-free reference image,  and combined analysis of photometry from multiple surveys -- will benefit from a flexible framework to model time-domain imaging of transients. We describe a time-domain extension of the \scarlettwo{} scene modeling code for multi-epoch, multi-band, and multi-resolution imaging data to extract simultaneous transient and host galaxy models. \scarlettwo{} leverages the benefits of data-driven priors on galaxy morphology, is fully GPU compatible, and can jointly model multi-resolution data from ground and space-based surveys. We demonstrate the method on simulated LSST-like supernova imaging, low-resolution Zwicky Transient Facility imaging of tidal disruption events, and Hyper Suprime Cam imaging of variable AGN out to $z=4$ in the COSMOS fields. We show that \scarlettwo{} models provide accurate transient and host galaxy models as well as accurate measurement of host--transient spatial offsets, and demonstrate future applications to the search for `wandering' massive black holes. 
\end{abstract}




\end{frontmatter}




\section{Introduction}
\label{introduction}
The next generation of wide-field optical time-domain surveys such as the Legacy Survey of Space and Time at Vera C. Rubin observatory \citep[LSST;][]{ivezic+08} and the High Latitude Time Domain Survey planned for the \textit{Nancy Grace Roman Space Telescope} \citep[\textit{Roman};][]{Akeson+19} will provide enormous discovery potential for high-redshift and low-luminosity transient phenomena such as supernovae (SNe), tidal disruption events (TDEs), and gamma ray burst (GRB) afterglows. Prompt classification of transient and variable objects -- essential to various extragalactic science cases where fast spectroscopic follow-up is particularly informative -- depends on our ability to correctly identify the host galaxies of transients, measure the distance between transients and their host galaxy nuclei, and estimate photometric redshifts from the host galaxy spectral energy distributions (SEDs). This becomes particularly challenging at the depths of LSST, where 30-50\% of galaxies in single epoch LSST images will overlap other galaxies, and 15\% of sources will be blended with other sources without being recognized as such \citep{Melchior2021TheSurveys}.

Time-domain surveys frequently employ methods based on a technique called `difference imaging' to detect transient phenomena. In such an approach, a deep `reference image' is produced from a stack of high quality images and subtracted from new images to produce a difference image \citep{Alard1998}. Difference imaging enables the efficient detection  and optimal extraction of multi-epoch fluxes of transient sources. In this work, we are primarily concerned with enabling follow-up investigations of the transient and its host once the transient has been identified by difference imaging. We propose a model of a variable point source \textit{and} a non-varying host galaxy that is rendered to match all observations. We refer to this as a \textit{scene modeling} approach, whereby all sources in the scene are modeled in a framework that incorporates information about observation parameters such as the world coordinate system, the point spread function, and the background noise. Our approach is particularly advantageous when the difference imaging introduces ambiguities -- for example, when the variable object is already contributing to the reference image, which is always the case for active galaxy nuclei (AGNs) -- or when simultaneously modeling the transient and the host is helpful, such as when measuring small spatial offsets between the host center and the transient. Our approach can also make use of imaging data from different surveys, enabling the automatic production of combined light curves without concerns about mismatches between the reference image epochs. 

Multi-resolution scene modeling methods have previously been applied to produce galaxy catalogs from deep stacks for the DESI Legacy Imaging Surveys \citep{Dey2019OverviewSurveys}, and to produce deep forced photometry of 400 million sources in low resolution imaging from the \textit{Wide-Field Infrared Survey Explorer}, exploiting higher resolution SDSS imaging to anchor the photometry \citep{Lang2016}. However, scene modeling has rarely been implemented for multi-epoch imaging of transients. A version of the scene modeling approach was applied to produce supernova photometry in the Sloan Digital Sky Survey-II Supernova survey \citep{Holtzman2008} and the Dark Energy Survey Supernova Program survey \citep{Brout2019}. The method is yet to be applied for nuclear transients and AGN, where degeneracies between emission from the galaxy nucleus and the AGN make deblending particularly challenging. Current best practice is to obtain a single epoch of high-resolution space-based imaging, and then to fit a galaxy profile and a nuclear point source to that image \citep[e.g.][]{Kim2008,Zhong2022AField}. In such cases, a prior on the AGN SED can be obtained by fitting AGN and galaxy templates to a spectrum of the galaxy nucleus. Convolutional neural networks have also been used for AGN-host decomposition of {\it James Webb Space Telescope} imaging \citep{Margalef2024}. To our knowledge, forward modeling of variable AGN across multi-epoch ground-based imaging has yet to be used as a method for AGN-host decomposition or extraction of AGN light curves.

Scene modeling is particularly powerful when imaging data from ground and space-based telescopes is combined so that the high-resolution imaging with small PSF sizes can improve galaxy deblending solutions for sources that are marginally resolved in low-resolution ground-based imaging \citep{Melchior2021TheSurveys}. Software packages that can simultaneously model multi-band, multi-resolution imaging data include \tractor{} \citep{2016ascl.soft04008L}, \scarlet{} 
 \citep{Melchior2018Scarlet:Factorization}, and \astrophot{} \citep{Stone2023scpastrophot/scpImages}, the latter of which is GPU-accelerated. In contrast to \tractor{} and \astrophot, \scarlet{} is a non-parametric method: it does not require the adoption of an analytical profile such as a S\'{e}rsic profile for extended sources. Simple requirements on galaxy shape, such as a monotonically decreasing or symmetric profile, or positive flux in all regions of the galaxy, are enforced by proximal operators. Gradient descent with \texttt{Autograd} is used to fit the source parameters, with the differing colors of overlapping sources being the most important distinguishing factor for deblending. The assumption that the morphology is constant across bands provides a strong constraint that is essential when modeling crowded galaxy fields. For large galaxies with color gradients, multi-component galaxy models enable accurate deblending while working under this constraint.

Recently, a new version of \scarlet{} was introduced (\citet{Sampson2024}; Melchior et al., in prep.). \scarlettwo{} is built with \texttt{jax}\ \citep{jax2018} and \texttt{equinox} \citep{equinox2021} for CPU/GPU compatibility. Importantly, \scarlettwo{} has the ability to utilize data-driven priors for galaxy morphologies in the form of score-based neural networks, which replace the hard physical constraints placed on the morphology models in \scarlet. The galaxy morphology priors significantly stabilize non-parametric model fitting of low-S/N and highly blended sources. In the context of this work, flexible galaxy models are essential for producing transient photometry via scene modeling, as the enforcement of a parametric galaxy profile that does not fully describe the data will generate seeing-dependent errors on the measured transient flux. \scarlettwo{} also supports \texttt{numpyro} \citep{Phan2019, Bingham2019} to undertake MCMC sampling over source parameters to enable careful quantification of uncertainties in source positions, SEDs and morphologies. The combination of GPU acceleration for fast fitting, MCMC sampling, and the capacity to fit non-parametric galaxy models makes \scarlettwo{} a capable base for a wide range of scene-fitting applications.

In this paper, we introduce a new time-domain extension to \scarlettwo{} that enables modeling of multi-band, multi-resolution \textit{and} multi-epoch imaging data, where we treat some sources as variable and others as static. We test the methodology on simulated multi-epoch, multi-band supernova imaging, and then demonstrate it on real survey data for samples of a) TDEs and b) AGNs. For our two case studies, we apply our scene-modeling technique to a) high cadence, low resolution, shallow multi-epoch imaging from the Zwicky Transient Facility \citep[ZTF;][]{Graham2019,Bellm2019TheResults,Dekany2020TheSystem} and b) low cadence, high resolution, deep multi-epoch imaging from the Hyper Suprime Cam Subaru Strategic Program (HSC-SSP) Transient Survey \citep{Yasuda2019TheOverview}. We demonstrate the application of non-parametric galaxy morphology models, guided by data-driven neural network priors, to produce full scene models with accurate transient photometry and position measurements. 


\section{Modeling variable and static sources in multi-epoch imaging}
\label{sec:modeling}

\begin{figure*}[h]
\begin{subfigure}{.5\textwidth}
  \centering
  \includegraphics[width=.98\linewidth]{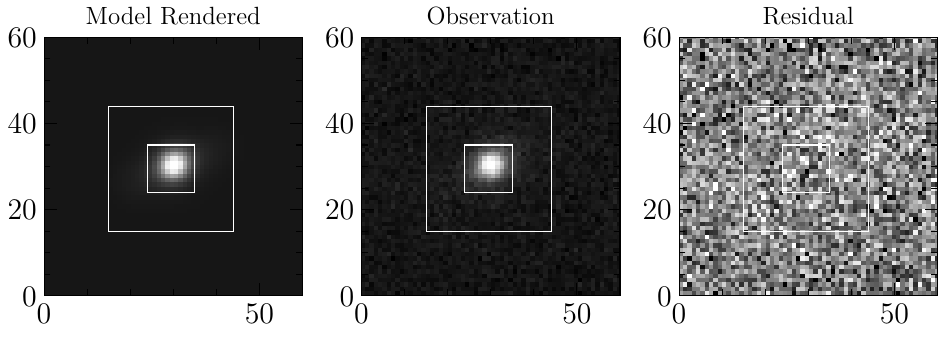}  
  \caption{Nuclear transient in bright state}
\end{subfigure}
\begin{subfigure}{.5\textwidth}
  \centering
  \includegraphics[width=.98\linewidth]{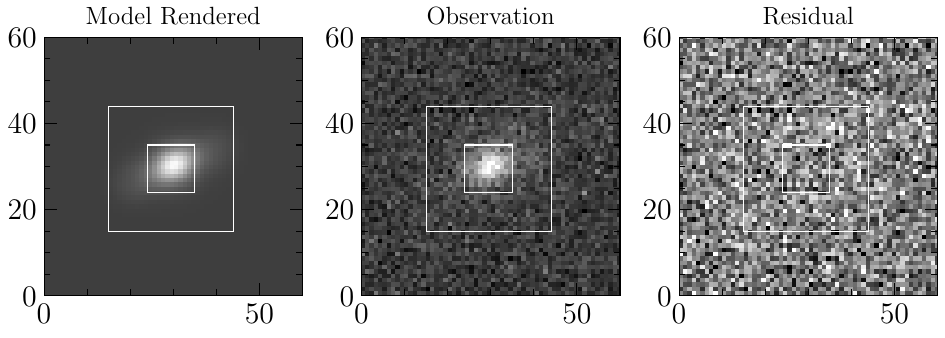}  
  \caption{Nuclear transient in faint state}
\end{subfigure}
\begin{subfigure}{.5\textwidth}
  \centering
  \includegraphics[width=.98\linewidth]{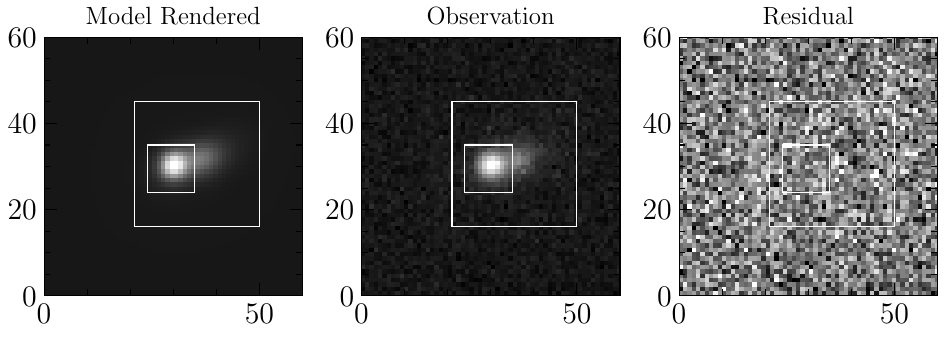}  
  \caption{Offset transient in bright state}
  \label{fig:sub-second}
\end{subfigure}
\begin{subfigure}{.5\textwidth}
  \centering
  \includegraphics[width=.98\linewidth]{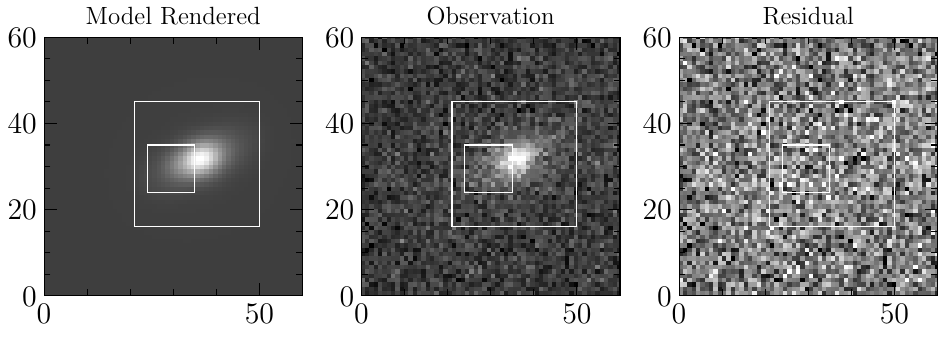}  
  \caption{Offset transient in faint state}
  \label{fig:sub-second}
\end{subfigure}
\caption{Example of a simulated supernova and host galaxy system and the resulting \scarlettwo{} models for two different host-transient spatial offsets. \scarlettwo{} models were derived from simulated imaging with a 0.17" pixel scale over 10 epochs and 3 bands ($g$, $r$, and $i$). \textit{Top}: A transient coincident with the galaxy nucleus while it is bright relative to the host galaxy (left) and faint relative to the host galaxy (right). \textit{Bottom:} Same as above but for a transient offset by 1.06".}
\label{fig:galsimexamples}
\end{figure*}

In \scarlet{}, a multi-band image cube $\mathbf{Y}$ is described as the linear combination of $K$ sources, each of which with its own amplitude $\mathbf{A}_k$ and intensity variation (morphology) $\mathbf{S}_k$:
\begin{equation}
\mathbf{Y} = \sum_k^K \mathbf{A}_k \times \mathbf{S}_k.
\end{equation}
The amplitude vector $\mathbf{A}_k\in\mathbb{R}^B$ has dimension equal to the number of different bands $B$ and contains the source spectrum or SED.
The source morphology image $\mathbf{S}_k\in\mathbb{R}^{HxW}$ has a size given by the height and width of either the observations $\mathbf{Y}$ or the area of the sky we seek to model.
This construction allows each source to have a free-form morphology that is constant across bands up to a multiplicative amplitude rescaling.

We now extend this definition for time-domain models, by generalizing $\mathbf{A}_k\in\mathbb{R}^E$ to contain the amplitudes for the total number of observed images $E$, which can be different in wavelength \emph{and in time}.
We assume to know the presence of static sources from regular source detection, and of variable sources from difference imaging, and we use this information to create a scene containing variable sources (which are allowed to change in flux across epochs) and `static' sources (which must not change their spectrum across epochs). 
For the former, the amplitude is independent across all epochs: 
\begin{equation}
\label{eq:A_variable}
\mathbf{A}_k^\mathrm{variable} = \{A_{k, e} \text{ for } e \in E\}.
\end{equation}
For the latter, we construct the full amplitude vector
\begin{equation}
\label{eq:A_static}
\mathbf{A}_k^\mathrm{static} = \{\bar{A}_{k, b(e)} \text{ for } e \in E\}
\end{equation}
from a time-independent SED vector $\bar{\mathbf{A}}_k\in\mathbb{R}^B$, where $b(e)$ refers to the band used for exposure $e$.

We also enable the user to optionally incorporate information about the exposures $E_{\text{on}}$ when the transient is present and may therefore have a non-zero flux and model the transient amplitude as
\begin{equation}
\label{eq:A_on}
\mathbf{A}_{k,e}^\mathrm{variable} =  \begin{cases} 
      A_{k,e} & \text{ if } e \in E_{\text{on}}\\
      0 & \text{else}.\\
   \end{cases}
\end{equation}

By providing \scarlettwo{} with information about the epoch and band of each image, and then defining sources as having either a variable or a static spectrum, a scene model with both varying and unvarying sources may be fit across multi-band, multi-epoch imaging data. Additional information about when a transient is `on' or `off' can be used as an additional constraint for transients with pre- or post- flare imaging, helping images from the transient-free epochs to improve the host galaxy SED and morphology model.

\section{Photometric and astrometric quality in simulated supernova imaging}
\label{sec:sne}

\begin{figure*}
\centering
\begin{subfigure}{.46\textwidth}
  \centering
  \includegraphics[width=.8\linewidth]{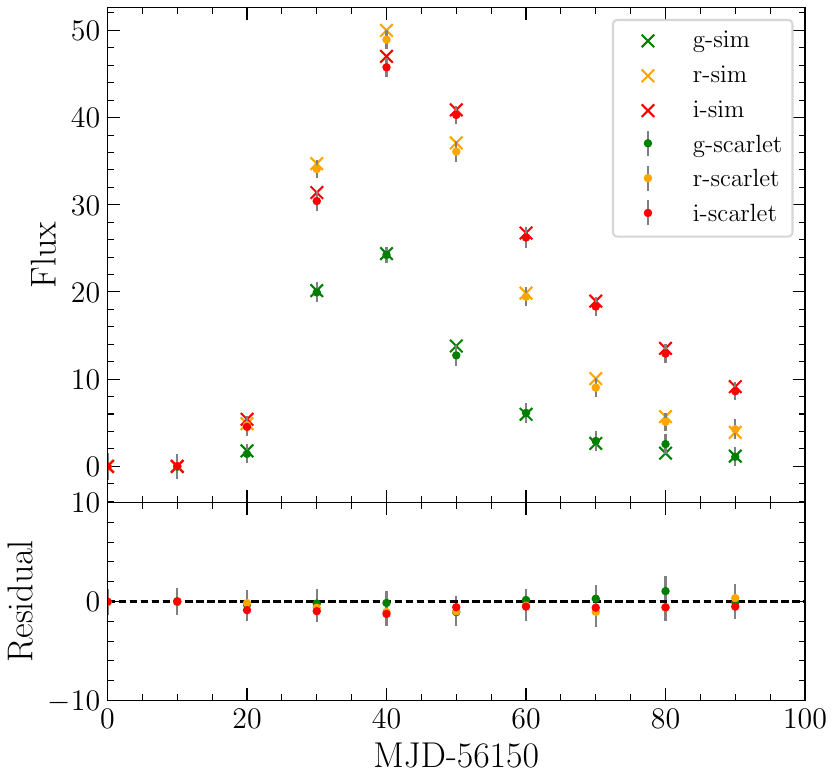}  
  \caption{\centering1.06" offset supernova with non-parametric host model and transient on/off constraint}
\end{subfigure}
\begin{subfigure}{.46\textwidth}
  \centering
  \includegraphics[width=.8\linewidth]{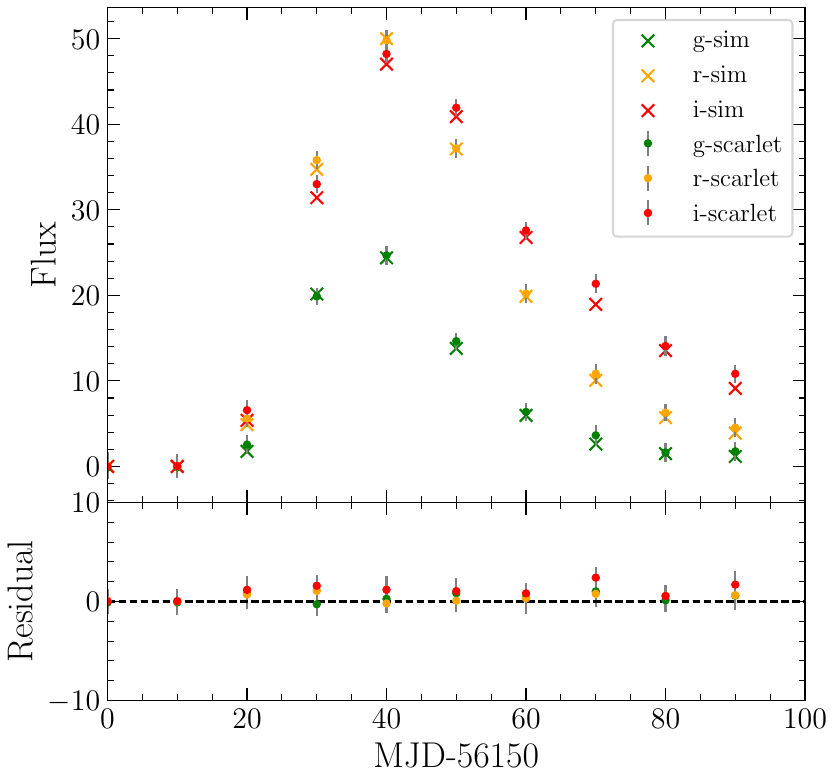}  
  \caption{\centering Nuclear supernova with non-parametric host model and transient on/off constraint}
\end{subfigure}
\begin{subfigure}{.46\textwidth}
  \centering
  \includegraphics[width=.8\linewidth]{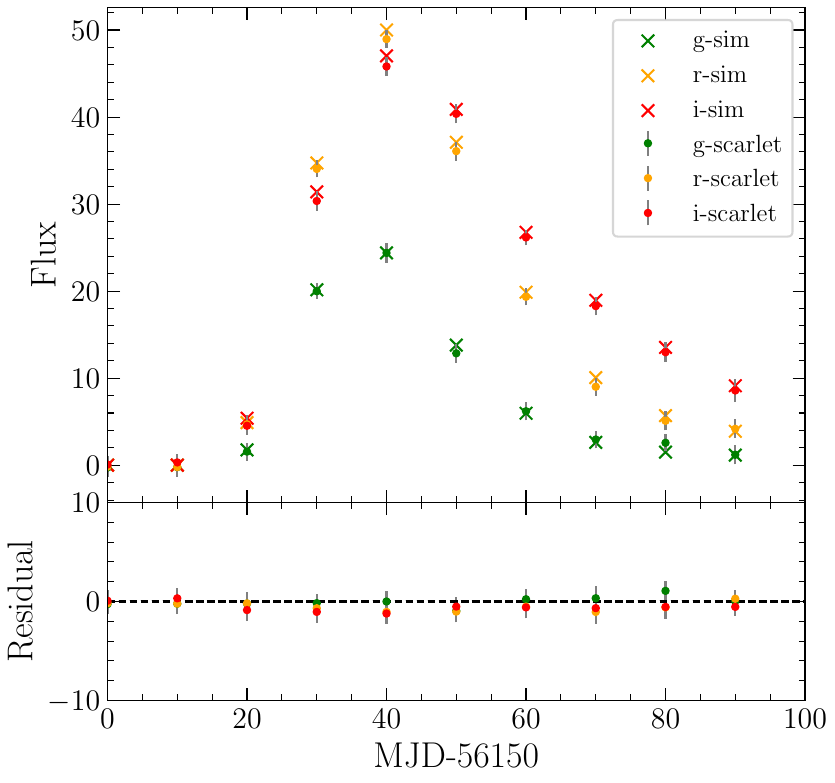}  
  \caption{\centering 1.06" offset supernova with non-parametric host model without transient on/off constraint}
\end{subfigure}
\begin{subfigure}{.46\textwidth}
  \centering
  \includegraphics[width=.8\linewidth]{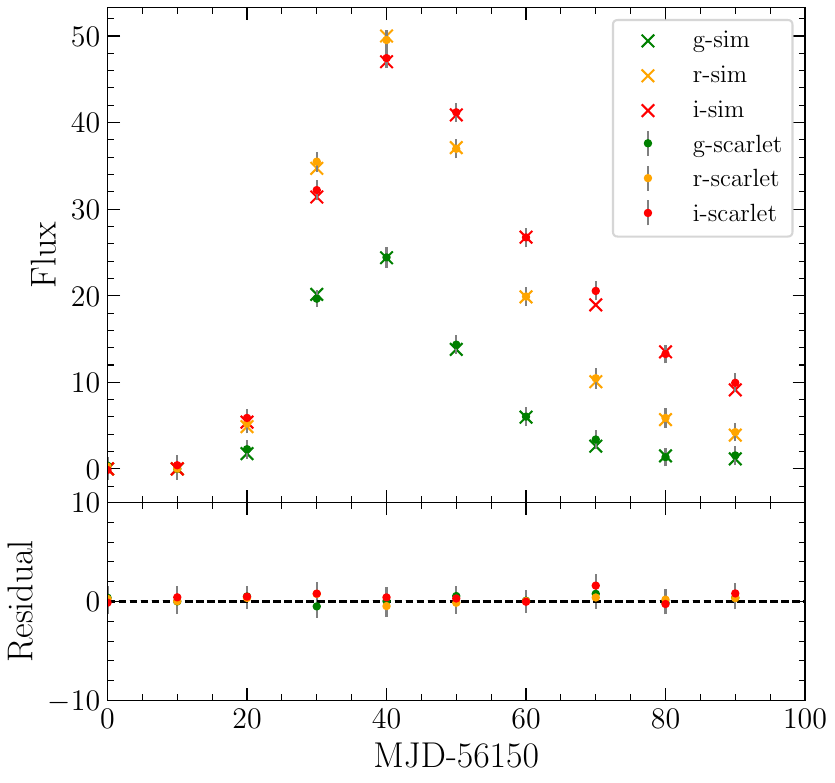}
  \caption{\centering Nuclear supernova with non-parametric host model without transient on/off constraint}
  \end{subfigure}
  \begin{subfigure}{.46\textwidth}
  \centering
  \includegraphics[width=.8\linewidth]{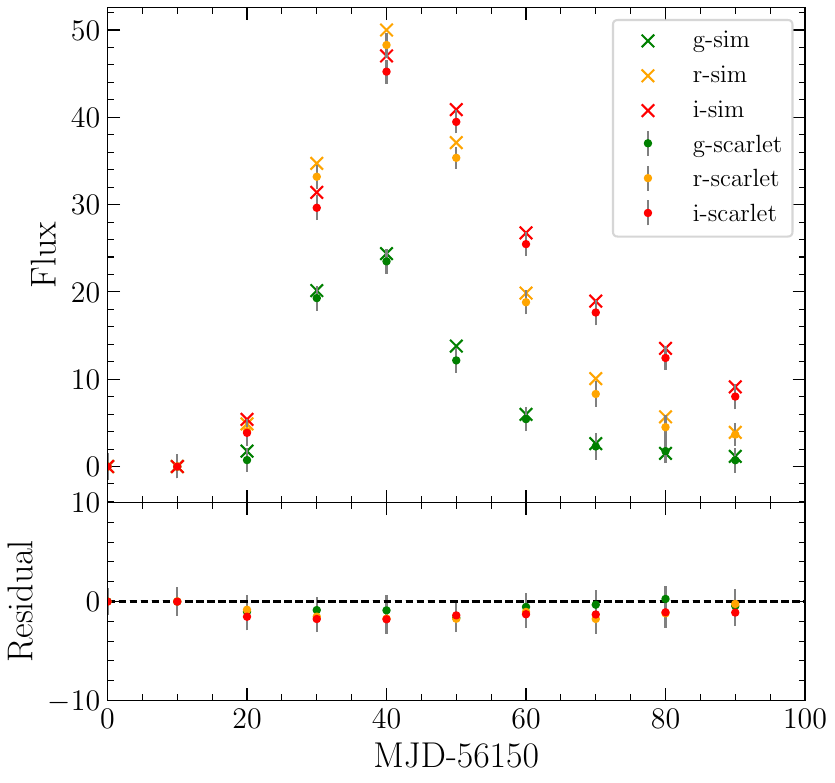}  
  \caption{\centering 1.06" offset supernova with S\'{e}rsic host model}
\end{subfigure}
\begin{subfigure}{.46\textwidth}
  \centering
  \includegraphics[width=.8\linewidth]{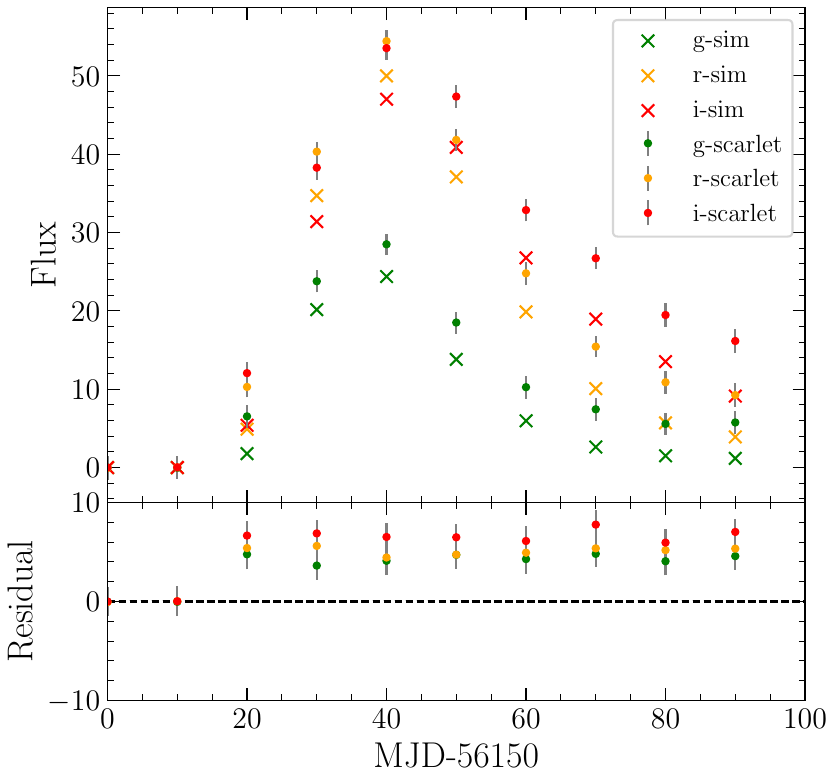}  
  \caption{\centering Nuclear supernova with S\'{e}rsic host model}
\end{subfigure}
\caption{\scarlettwo{} light curves compared to simulated values for a supernova located  1.06" from the galaxy nucleus (left) and in the galaxy nucleus (right). We show $3\sigma$ uncertainties from the posterior derived from MCMC sampling over the supernova flux and position, and host galaxy SED and morphology. We show results from modeling with three different constraints: A non-parametric host galaxy model with a transient that is constrained as `on' during the supernova and `off' otherwise (top row), a non-parametric host galaxy model with a transient that has no on/off constraint applied (middle row), and a  S\'{e}rsic host model with a transient that has the on/off constraint applied (bottom row). The flexible non-parametric morphology model does not impose an analytical profile onto realistic galaxies that may introduce biases in the transient photometry.}
\label{fig:lightcurves}
\end{figure*}

With the extension from \autoref{sec:modeling}, \scarlettwo{} can produce light curves and measure transient--host spatial offsets for a suite of simulated supernovae in LSST-like multi-epoch imaging. To produce the multi-epoch images, we first used \texttt{SNCosmo} \citep{kyle_barbary_2016_168220} to simulate the multi-band light curve of an SNIa at redshift $z=0.4$, generating $g$, $r$ and $i$ band fluxes for 15 epochs over a period of 240 days beginning 100 days before the onset of the supernova. We used \texttt{galsim} \citep{galsim2015} to simulate realistic images of the supernova and its host galaxy. We generated two sets of simulated images: one for an extended host galaxy with a bulge radius of 0.8" and a disk radius of 1.8", and one for a compact host galaxy with a bulge radius of 0.3" and a disk radius of 0.5". In each case the galaxy axis ratio was set to 0.43 and the position angle to 23 degrees. Images of size 65 by 65 pixels, with a pixel scale of 0.17"/pixel, a PSF FWHM of 0.8", and a sky noise level of 10 ADU per arcsec$^2$, were generated for each band and epoch. We generated the set of 45 multi-band, multi-epoch images for a series of spatial offsets between the supernova and the host center: 0.0", 0.03", 0.07", 0.14", 0.28", 0.49", 0.71", 1.06", 1.41", and 2.12".

To generate \scarlettwo{} scene models of the simulated supernova images, we first ran the wavelet detection routine implemented in \scarlet \, \citep{Melchior2018Scarlet:Factorization} on the summed images using the first 3 wavelet levels. If a single source was detected, we initialized the \scarlettwo{} model with a point source and a single extended source at the position found by the detection routine. If the detection routine found 2 sources (which was the case for offsets $>0.5$"), we initialized the point source and extended source models at the best-fit positions found by the detection routine. 

The extended sources were modeled with the non-parametric \texttt{Source} class,  with initial morphologies taken from an elliptical Gaussian fit to the summed image and a \texttt{StaticArraySpectrum}, following \autoref{eq:A_static}, with initial values obtained from the peak pixel. The Source morphology was given the positive constraint and the \texttt{ZTF\_ScoreNet32} prior from the score-model of the \texttt{galaxygrad}\footnote{\url{https://github.com/SampsonML/galaxygrad}} package \citep{Sampson2024}.
This model was trained to match the distributions of extended source galaxies in ZTF, which will match the host properties of our simulations well overall but lack finer detail.

The SNe are modeled as {\texttt PointSource} objects, i.e. their morphology is identical to the PSF. Their spectra were either complete free \texttt{ArraySpectrum}, following \autoref{eq:A_variable}, or \texttt{TransientArraySpectrum}, following \autoref{eq:A_on}, depending on whether the transient on/off constraints were being tested, with initial values taken from the pixel associated with their detection location. \scarlettwo{} then fit the scene until a relative error of $10^{-6}$ was reached, or a maximum of 3000 steps.

After obtaining the scene model, we used the \texttt{numpyro} NUTS MCMC sampling routine \citep{Phan2019, Bingham2019} implemented within \scarlettwo{} to sample over the point source position, the host galaxy spectrum, and the supernova flux in each epoch and estimate their uncertainties. For this exploratory study, we set a Gaussian prior of width given by 10\% of the best-fit value for each parameter. We ran the MCMC sampler for 200 burn-in iterations and 500 sampling iterations.

In \autoref{fig:galsimexamples} we show observations, rendered models, and residuals based on the results of the full multi-epoch scene model for one $g$-band image when the supernova is bright and one pre-flare image.
We distinguish two cases: In the top row, the supernova is located at the center of the galaxy nucleus; in the bottom row, it has a 1.06" offset from the galaxy nucleus. By modeling the full sets of 45 images with a variable point source and static host galaxy, \scarlettwo{} is able to produce accurate scene models for all single-epoch images. In \autoref{fig:lightcurves} we show the best-fit \scarlettwo{} point-source flux for each epoch vs the true simulated flux. \scarlettwo{} provides accurate photometry for all simulated supernovae, even when the supernova is coincident with the galaxy nucleus. We show light curves for simulated supernova imaging obtained both with and without the transient on/off constraint, and find that the inclusion of pre-supernova imaging is sufficient to accurately decompose the supernova and galaxy flux, regardless of whether the constraint is applied. 

In \autoref{fig:lightcurves}, we also compare the quality of the transient photometry derived when using a non-parametric galaxy morphology model to an alternative approach where \scarlettwo{} is required to fit a S\'{e}rsic galaxy model, with free parameters for the central position, ellipticity, half-light radius, and S\'{e}rsic index describing the steepness of the profile. As the selected parametric model cannot fully describe the  simulated disk and bulge galaxy profiles, the galaxy flux is underestimated in the center and overestimated in the outskirts, producing underestimated transient fluxes for the spatially offset supernova and overestimated transient fluxes for the nuclear supernova. While a more complex multi-component parametric model could improve the transient photometry in this case, the non-parametric model provides a flexible approach that enables accurate transient photometry without careful selection of the best parametric model.

\begin{figure}[h]
	\centering 	\includegraphics[width=0.45\textwidth]{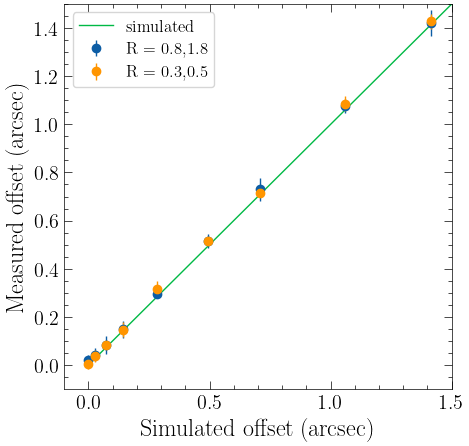}	
	\caption{Measured transient--host spatial offsets from simulated imaging of a transient at various distances from the nucleus of its host galaxy. We show the results for two host galaxies of different sizes: one with bulge radius of 0.8" and disk radius of 1.8" (corresponding to the images in Figure \ref{fig:galsimexamples}) shown in blue, and another with a bulge radius of 0.3" and disk radius of 0.5" shown in orange. Uncertainties are $1\sigma$ uncertainties derived from the posterior that was obtained when sampling over the source parameters.} 
	\label{fig:galsimoffsets}%
\end{figure} 

In order to calculate the spatial offsets between the simulated supernovae and their host galaxy, we first needed to determine the center of the galaxy nucleus to sub-pixel accuracy. To do this, we first prepared a high S/N galaxy image by subtracting the rendered \scarlettwo{} model of the transient from each observation, and then stacking those transient-subtracted multi-epoch images to produce a high S/N multi-band image of the host galaxy. We then fit a S\'{e}rsic galaxy model to the transient-subtracted image to fit the center of the galaxy by initializing a S\'{e}rsic profile and allowing \scarlettwo{} to fit the half light radius, ellipticity, S\'{e}rsic index, spectrum, and central position. The MCMC sampling routine was again applied to determine the uncertainties for the galaxy model parameters. The offset between the best-fit point source position derived from the original non-parametric scene model and the best-fit galaxy center from the S\'{e}rsic profile fit to the transient-subtracted image stack was calculated for each set of simulated images. In \autoref{fig:galsimoffsets} we show the best-fit spatial offsets found by the \scarlettwo{} fits vs the known simulated offset for the compact galaxy profile and the extended galaxy profile. Our fitting routine is able to correctly measure both small and large spatial offsets between supernovae and their host galaxies, for both extended and compact host morphologies.

\section{Tidal disruption events in low-resolution Zwicky Transient Facility imaging}
\subsection{Background and motivation}

\begin{figure*}[h]
\begin{subfigure}{.98\textwidth}
  \centering
  \includegraphics[width=.8\linewidth]{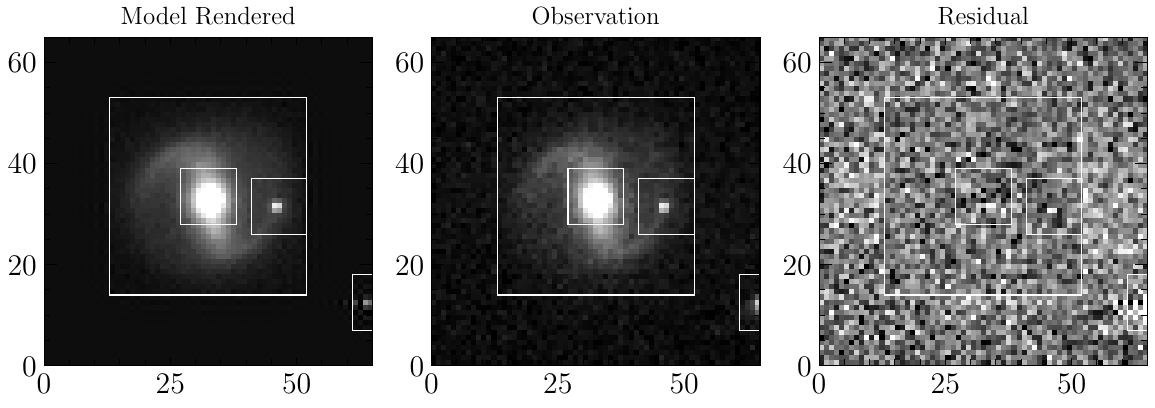}  
  \caption{ZTF19abzrhgq}
\end{subfigure}
\begin{subfigure}{.98\textwidth}
  \centering
  \includegraphics[width=.8\linewidth]{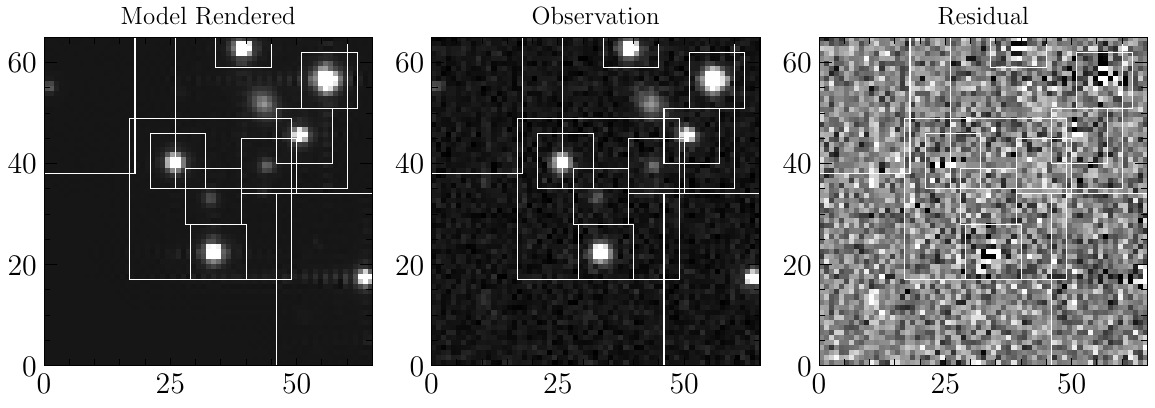}  
  \caption{ZTF19abhhjcc}
\end{subfigure}
\caption{Example \scarlettwo{} rendered models, observations, and residuals produced by modeling the full multi-epoch, multi-band ZTF imaging dataset for two TDEs. We show a TDE with a large and complex host galaxy above and a TDE with a compact host galaxy in a busy field below. Some artifacts due to the undersampled PSF are visible.}
\label{fig:TDEeg}
\end{figure*}

\begin{figure*}[h]
\begin{subfigure}{.48\textwidth}
  \centering
  \includegraphics[width=.98\linewidth]{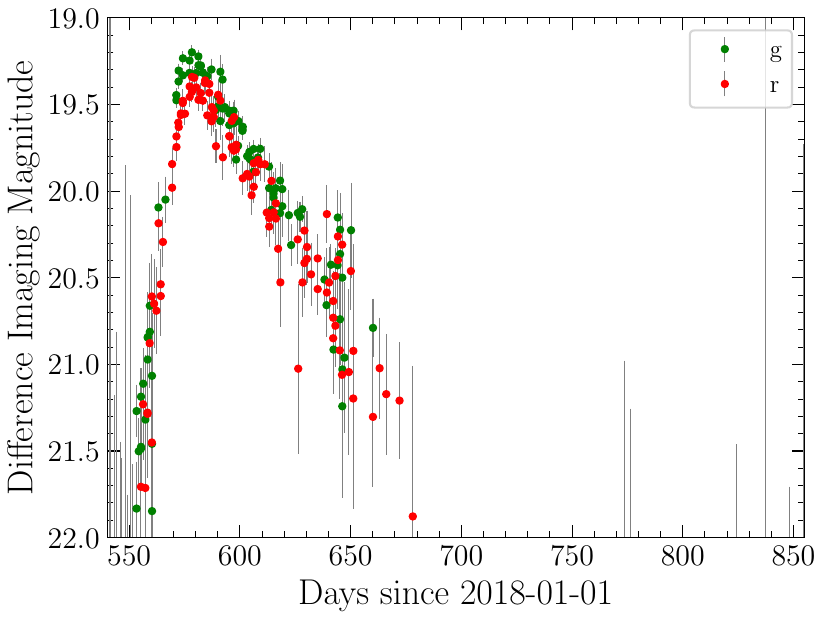}  
  \caption{ZTF difference imaging light curve}
\end{subfigure}
\begin{subfigure}{.48\textwidth}
  \centering
  \includegraphics[width=.98\linewidth]{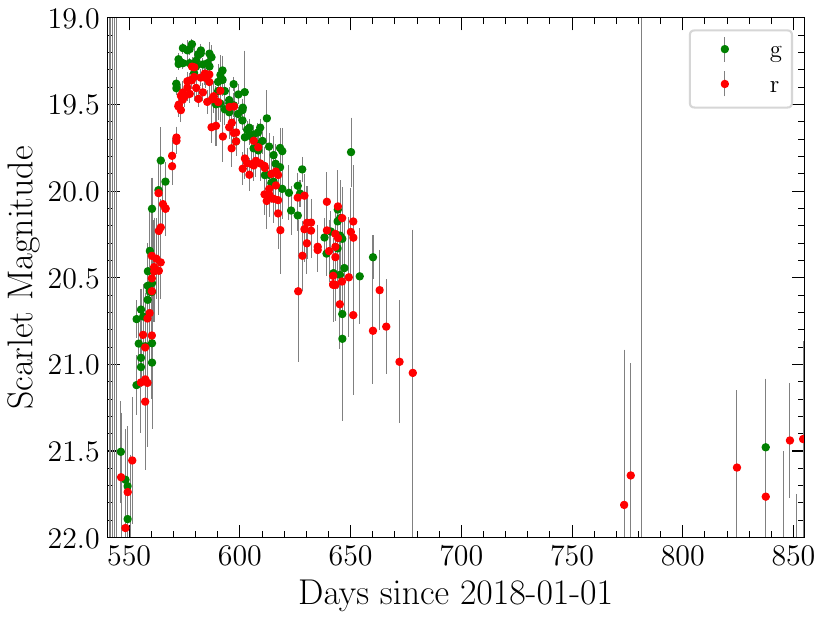}  
  \caption{\scarlettwo{} light curve}
\end{subfigure}
\caption{Comparison of light curves of example TDE ZTF19abhhjcc produced by forced photometry on ZTF difference imaging (left) and \scarlettwo{} scene modeling (right).}
\label{fig:TDElcs}
\end{figure*}
\begin{figure*}[h]
\centering
\begin{subfigure}{0.45\columnwidth}
  \centering
  \includegraphics[width=.99\linewidth]{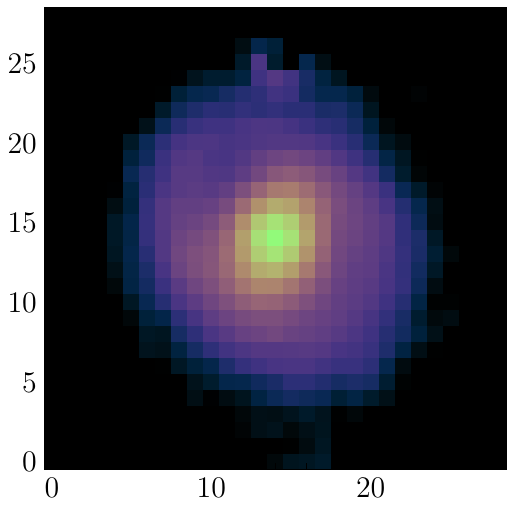}  
  \caption{\centering ZTF19aapreis}
\end{subfigure}
\begin{subfigure}{0.45\columnwidth}
  \centering
  \includegraphics[width=.99\linewidth]{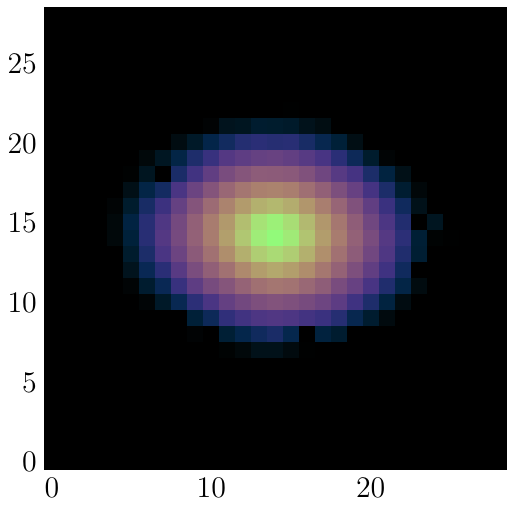}  
  \caption{\centering ZTF19aarioci}
  \end{subfigure}
  \begin{subfigure}{0.45\columnwidth}
  \centering
  \includegraphics[width=.99\linewidth]{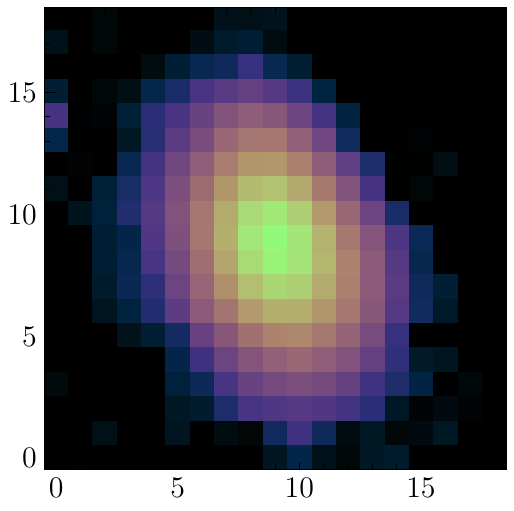}  
  \caption{\centering ZTF19accmaxo}
\end{subfigure}
\begin{subfigure}{0.45\columnwidth}
  \centering
  \includegraphics[width=.99\linewidth]{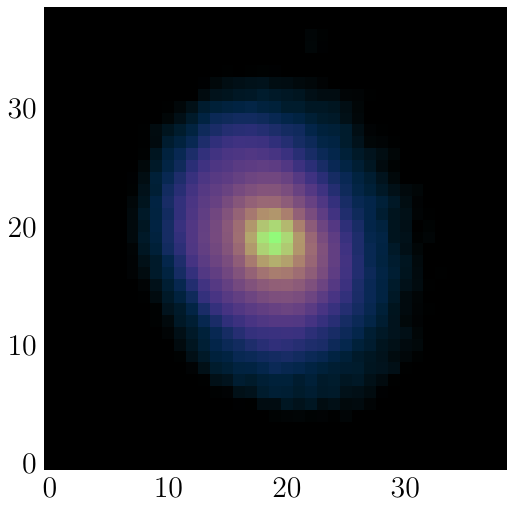}  
  \caption{\centering ZTF17aaazdba}
\end{subfigure}
  \begin{subfigure}{0.45\columnwidth}
  \centering
  \includegraphics[width=.99\linewidth]{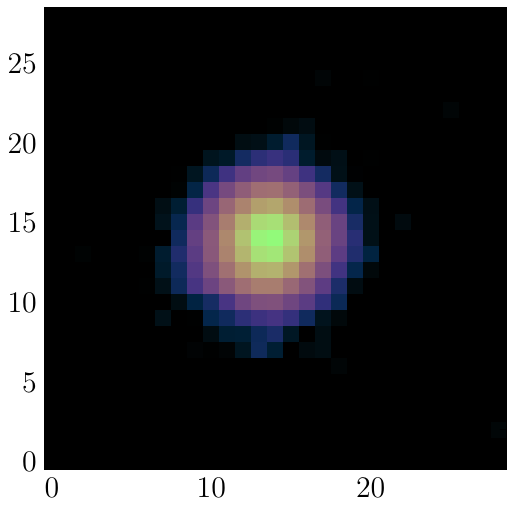}  
  \caption{\centering ZTF19aakswrb}
\end{subfigure}
\begin{subfigure}{0.45\columnwidth}
  \centering
  \includegraphics[width=.99\linewidth]{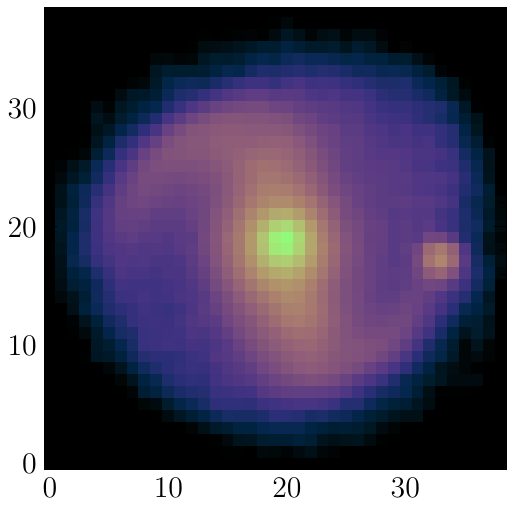}  
  \caption{\centering ZTF19abzrhgq}
\end{subfigure}
  \begin{subfigure}{0.45\columnwidth}
  \centering
  \includegraphics[width=.99\linewidth]{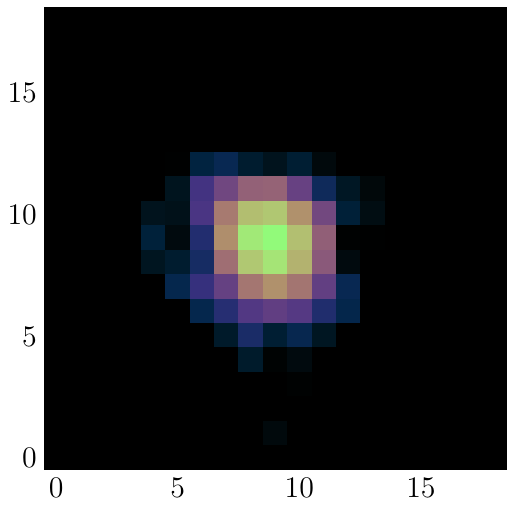}  
  \caption{\centering ZTF19abhejal}
\end{subfigure}
\begin{subfigure}{0.45\columnwidth}
  \centering
  \includegraphics[width=.99\linewidth]{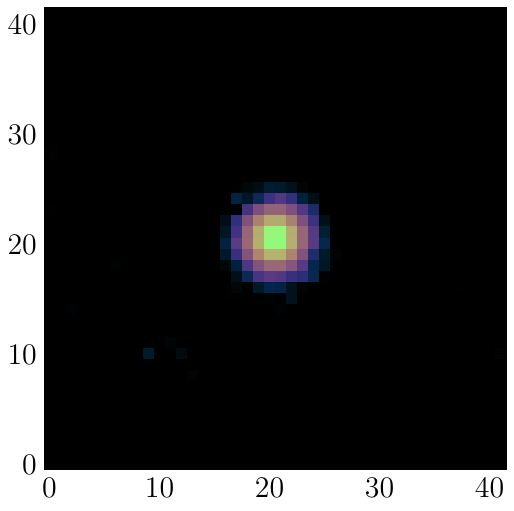}  
  \caption{\centering ZTF19aatylnl}
\end{subfigure}
\caption{Example \scarlettwo{} morphology models for 8 TDE host galaxies in ZTF imaging, shown with logarithmic scaling. The non-parametric modeling approach, guided by the custom ZTF prior, enables accurate descriptions of asymmetric and non-monotonically decreasing galaxy morphologies (e.g. ZTF19aapreis and ZTF19abzrhgq).} 
\label{fig:TDEhost}
\end{figure*}

\begin{figure*}[h]
\begin{subfigure}{\textwidth}
  \centering
  \includegraphics[width=0.85\linewidth]{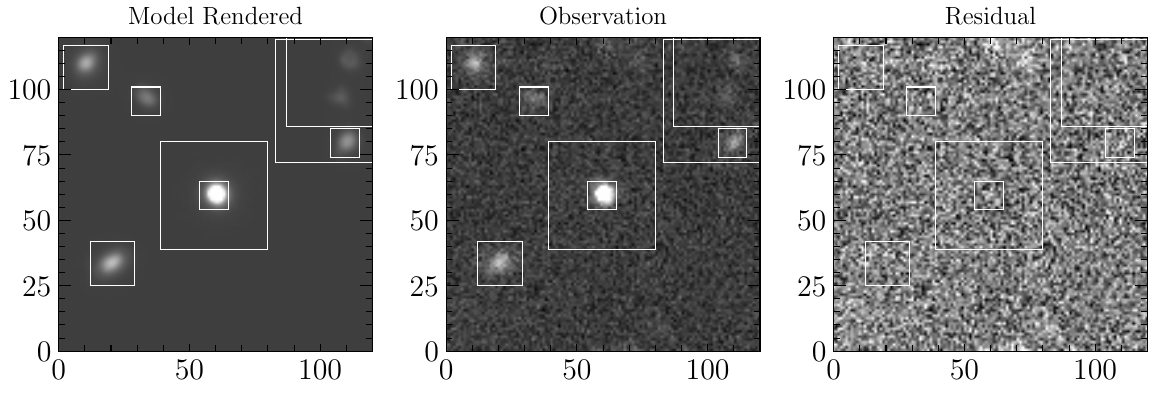}  
  \caption{COSMOS 834538}
\end{subfigure}
\begin{subfigure}{\textwidth}
  \centering
  \includegraphics[width=0.85\linewidth]{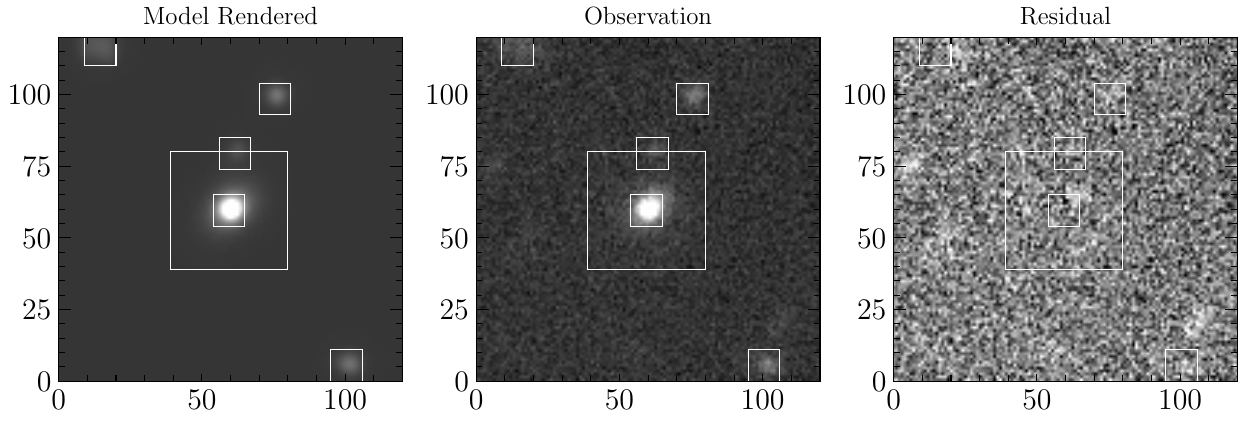}  
  \caption{COSMOS 505243}
\end{subfigure}
\caption{Example \scarlettwo{} models and residuals for single exposures based on modeling of the full multi-epoch, multi-band HSC imaging dataset for two variable COSMOS AGN. We show an AGN with a compact, faint host galaxy above and an AGN with faint but extended host galaxy emission below.}
\label{fig:HSCeg}
\end{figure*}

We extend our study to much lower resolution images and demonstrate how to use \scarlettwo{} to measure transient-host offsets and produce accurate photometry of low redshift transients in galaxy nuclei using the sample of 17 TDEs discovered in ZTF-I \citep{vanVelzen2021SeventeenStudies}. 
In the ZTF survey, measuring the distance between transients and their host nuclei was challenging given the 1.0" pixel scales. The `nuclear transient alert filter' for ZTF, which was designed to use difference imaging metadata to identify TDEs and AGN flares, required that the mean spatial offset between the position of the transient in all difference images containing detections and the centroid of the closest source in the reference image be less than 0.4" \citep{vanVelzen2021SeventeenStudies}. However, the distribution of spatial offsets measured for nuclear AGN via this method was shown to extend to $\lessapprox 0.95$" \citep{Ward2021AGNsFacility}. In one case, the optical counterpart to an X-ray identified TDE had an overestimated transient-host offset based on ZTF difference imaging statistics such that it did not meet the $<0.4"$ cutoff, and a full scene model of the point source and host galaxy in the ZTF imaging was required to ascertain that the optical transient seen in ZTF was indeed coincident with the galaxy nucleus and therefore associated with the X-ray TDE \citep{Brightman2021AEvent}. 


\subsection{Light curves, host galaxy morphology models, and spatial offsets for ZTF TDEs}

To produce \scarlettwo{} light curves and host galaxy models of the ZTF-I TDEs, we first used the \texttt{ZTFquery} cutout service \citep{Rigault2018ZtfqueryData} to download 120" by 120" cutouts of the ZTF single-epoch imaging of each TDE. We obtained imaging covering 900 days, beginning from 50 days prior to the TDE, and removed images with seeing FWHM$>2$" and limiting magnitude $<20$. This resulted in approximately 300 \textit{g}, \textit{r} and \textit{i} band images per scene. We repeated the procedure described for the simulated \texttt{galsim} imaging in \autoref{sec:sne}, running \scarlettwo{} with the 32 by 32 pixel ZTF prior to produce non-parametric models of each host galaxy and blended background galaxies, in addition to a variable point source initialized at the galaxy center. Example ZTF images and models of two TDEs are shown in \autoref{fig:TDEeg}. 

After obtaining a full scene model, including the single-epoch fluxes for the transient, we again produced a stack of the transient-subtracted single-epoch images to fit a S\'{e}rsic model to a high S/N host galaxy image to measure the center of the galaxy nucleus. From the MCMC sampling results, we find that all TDEs are consistent with the galaxy nuclei as expected, with all best-fit offsets $<0.9$". Typical $3\sigma$ spatial offset uncertainties reported from MCMC sampling over source SEDs, positions, and morphologies were 1-2", consistent with previous analysis of the spatial offset uncertainty distribution of variable AGN reported from ZTF difference imaging \citep{Ward2021AGNsFacility}.

In \autoref{fig:TDElcs} we show the the single-epoch fluxes measured by \scarlettwo{} for one of the ZTF TDEs, ZTF19abhhjcc, as well as the equivalent light curve produced using the forced photometry pipeline for ZTF difference imaging \citep{Masci2019}. \scarlettwo{} is able to produce a light curve consistent with the difference imaging pipeline, and correctly assigns flux to the host galaxy and the nuclear transient where they are blended. We specifically chose ZTF19abhhjcc because it represents a more difficult modeling problem: the host galaxy is compact (\autoref{fig:TDEeg}), such that there is little extended emission beyond the nucleus to assist in decomposing the nuclear transient and host galaxy SEDs. 

In \autoref{fig:TDEhost}, we show the non-parametric galaxy morphology models for 8 of the TDE host galaxies. The galaxy morphology models are able to describe complex shapes including spiral arms. By using a non-parametric model for the host galaxy and background galaxies we are able to produce high quality light curves that are not contaminated by residuals in the galaxy model, which can arise when enforcing a simplified parametric galaxy model. 

In summary, \scarlettwo{} scene modeling can correctly extract transient photometry, galaxy morphology models, and transient-host spatial offsets for our test sample of nuclear transients even in low-resolution ZTF imaging. Sampling over source parameters enables accurate quantification of transient--host spatial offsets and source SEDs. \scarlettwo{} can therefore be applied to large multi-epoch datasets to confirm whether TDEs and TDE candidates are coincident with galaxy nuclei.

\section{Variable AGN in the HSC-SSP Transient Survey}

\begin{figure*}[h]
\begin{subfigure}{.33\textwidth}
  \centering
  \includegraphics[width=\linewidth]{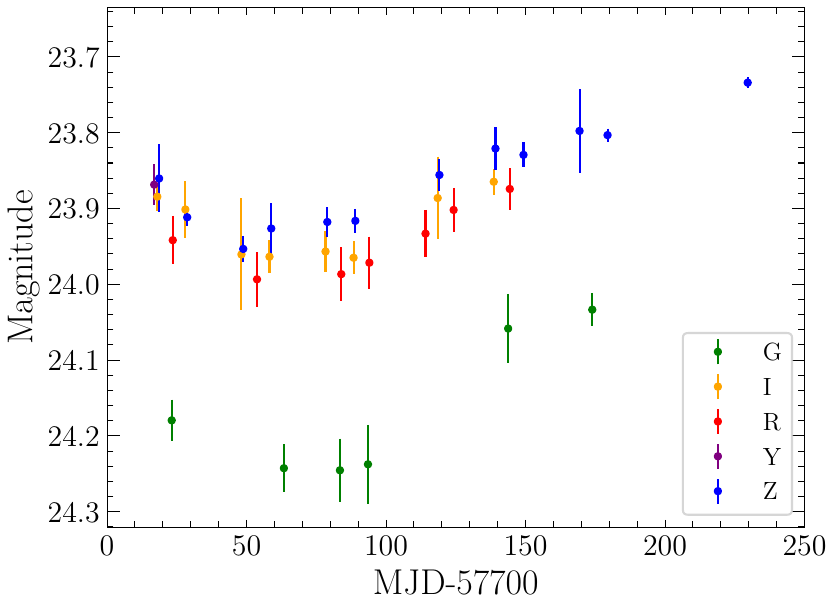} 
  \caption{\centering COSMOS 472392}
\end{subfigure}
\begin{subfigure}{.33\textwidth}
  \centering
  \includegraphics[width=\linewidth]{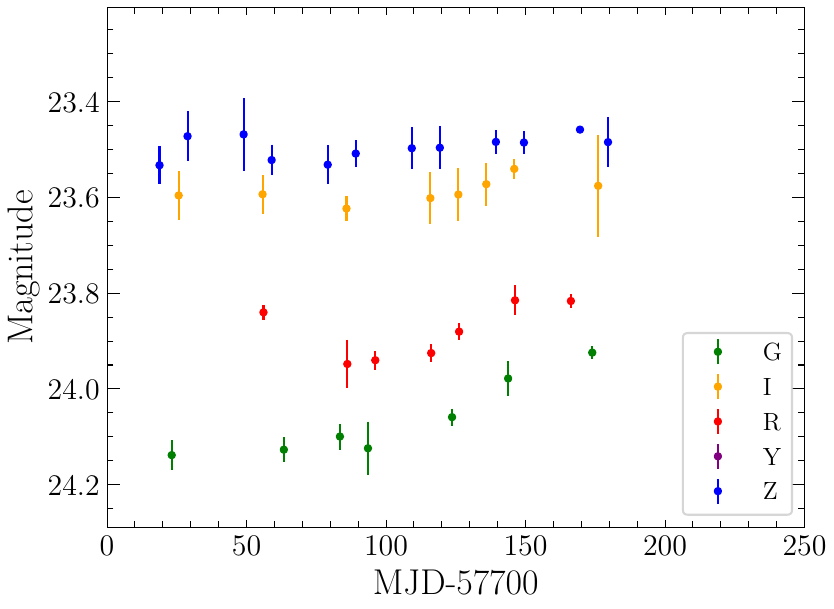}  
  \caption{\centering COSMOS 505243}
\end{subfigure}
\begin{subfigure}{.33\textwidth}
  \centering
  \includegraphics[width=\linewidth]{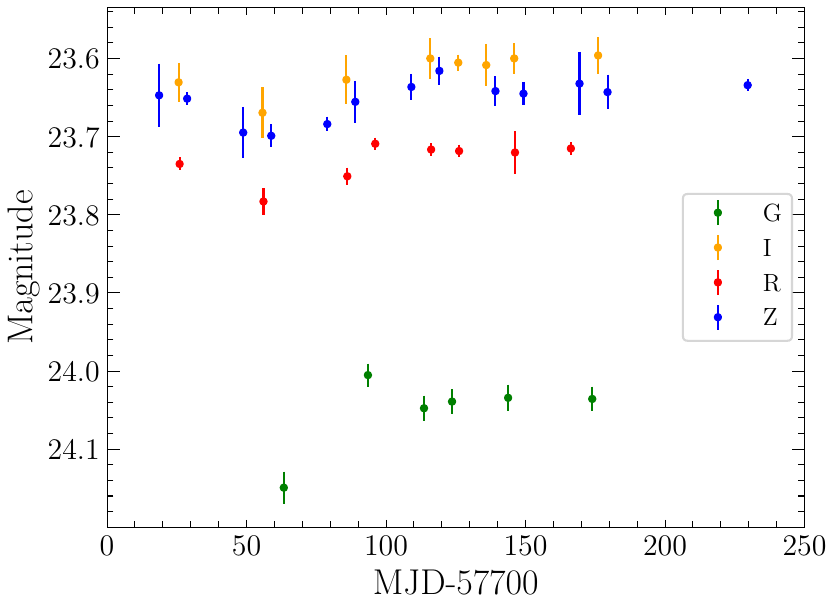}  
  \caption{\centering COSMOS 834538}
\end{subfigure}
\caption{Example \scarlettwo{} light curves of the AGN in the HSC-SSP transient survey, zoomed into a 250 day period with high cadence monitoring. Long term AGN variability at the $\sim0.1$ magnitude level consistent across 3 to 5 HSC bands is detected for all AGN in the sample.}
\label{fig:HSClc}
\end{figure*}

\begin{figure*}[h]
\begin{subfigure}{.49\textwidth}
  \centering
  \includegraphics[width=\linewidth]{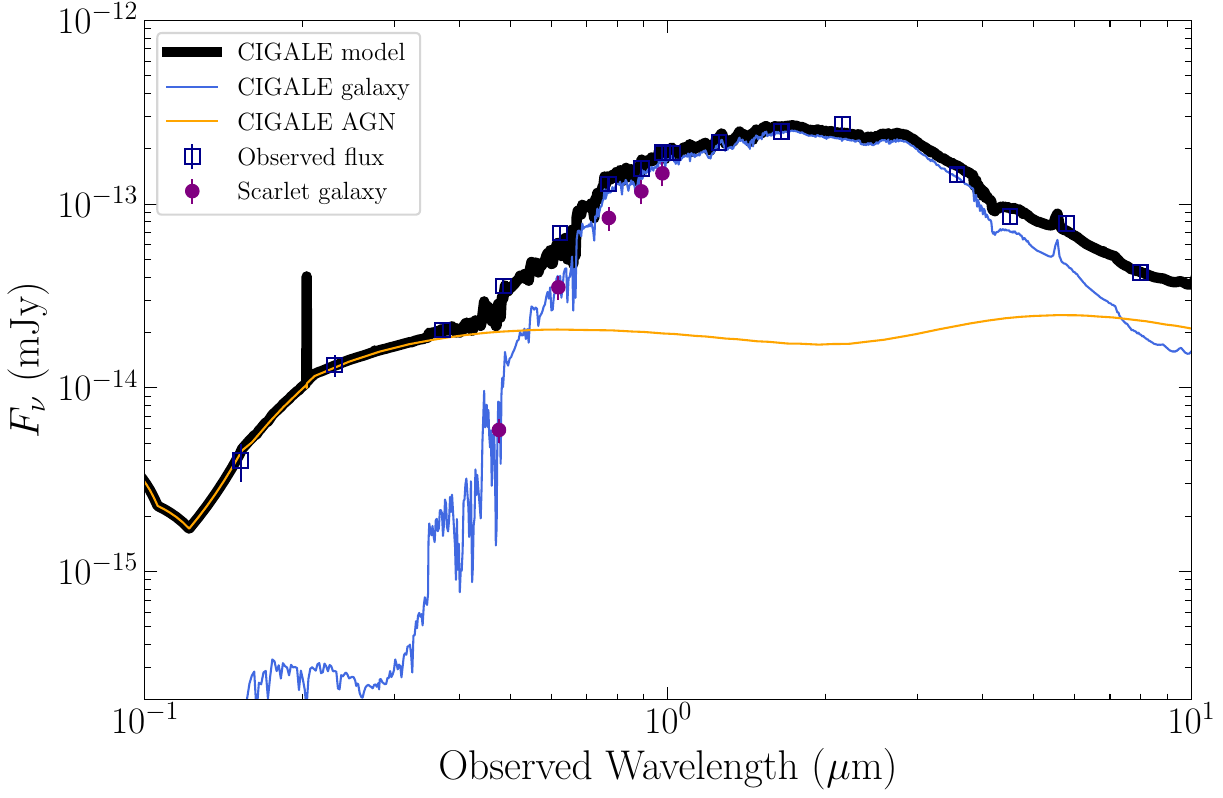} 
  \caption{\centering COSMOS 505243}
\end{subfigure}
\begin{subfigure}{.49\textwidth}
  \centering
  \includegraphics[width=\linewidth]{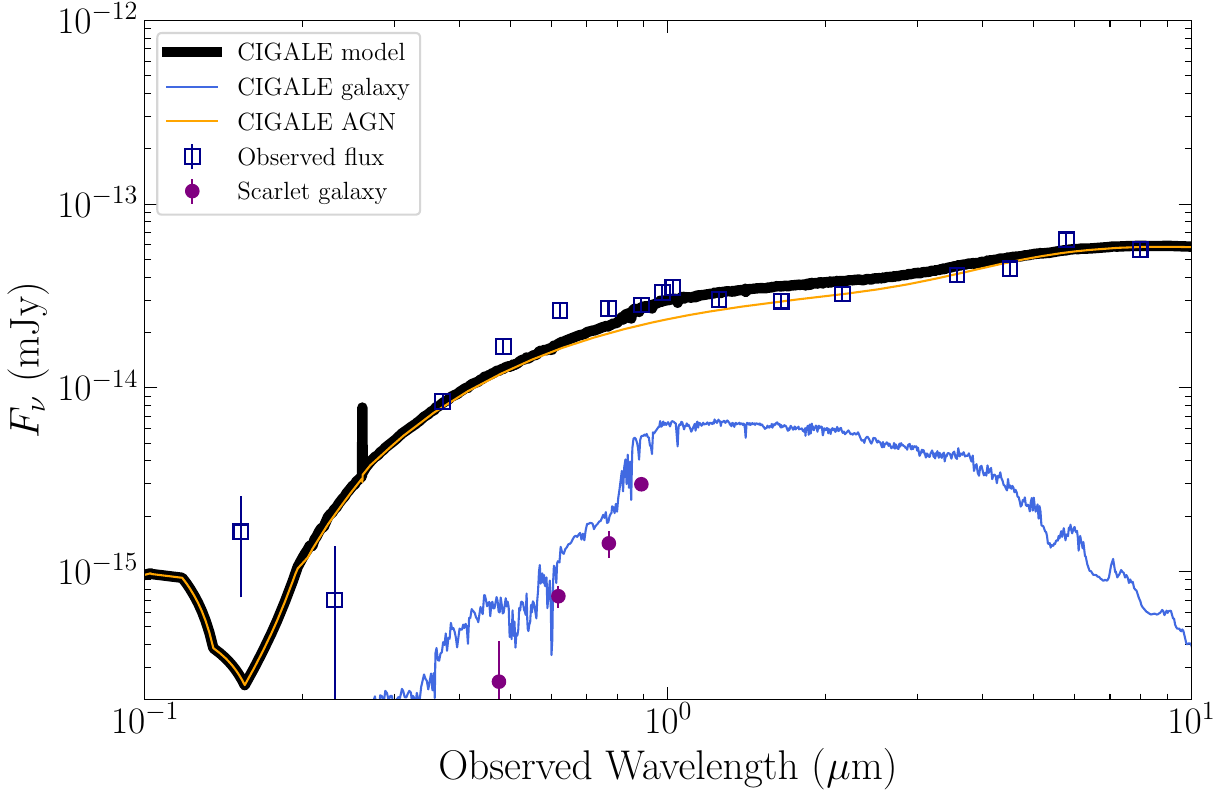}  
  \caption{\centering COSMOS 834538}
\end{subfigure}
\caption{Comparison of galaxy and AGN SEDs estimated from \texttt{cigale} modeling of combined host and AGN spectra and those estimated from multi-band, multi-epoch HSC image modeling with \scarlettwo. The observed fluxes derived from an optical spectrum are shown in blue boxes. The total \texttt{cigale} model derived from the fluxes is shown in black. The AGN contribution to the model is shown in orange and the host galaxy's stellar contribution in blue. Overlaid onto the \texttt{cigale} results are the best-fit scarlet SEDs for the host galaxy with 1$\sigma$ uncertainties derived from scene modeling of multi-epoch HSC imaging. We show the results for two galaxies: one where the \texttt{cigale} model estimates a substantial contribution from the host galaxy to the total flux in the wavelength range of the HSC imaging (left) and one where the \texttt{cigale} model estimates that the stellar component is dominated by the AGN component in the wavelength range of the HSC imaging (right). In both cases \scarlettwo{} finds results comparable to the spectroscopic decomposition via imaging modeling alone.}
\label{fig:SEDexamples}
\end{figure*}

\begin{figure}[h]
	\centering 	\includegraphics[width=0.49\textwidth]{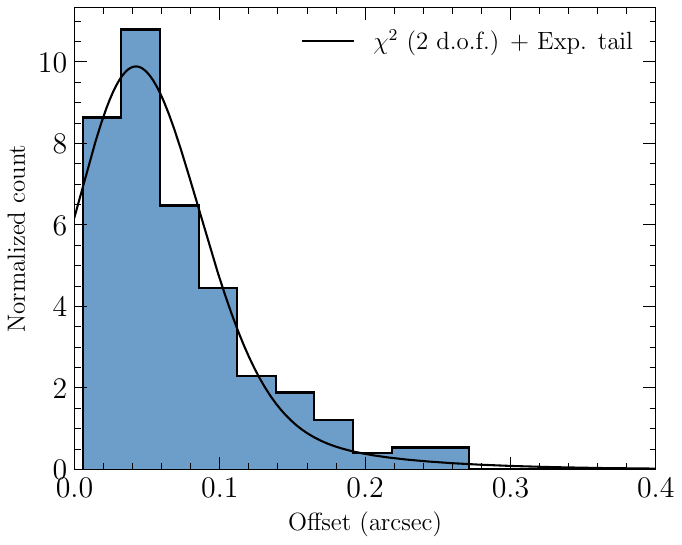}	
	\caption{Histogram of host-AGN offsets for the COSMOS AGN population derived from \scarlettwo{} modeling of the full multi-epoch, multi-band HSC-SSP transient survey imaging dataset. The best-fit model of a mixture distribution of a Rayleigh distribution of $\sigma = 0.051$" and an exponential tail with $\alpha=14.2$ is shown in black.} 
	\label{fig:HSCoffsets}%
\end{figure}

\begin{figure}[h]
	\centering 	\includegraphics[width=0.45\textwidth]{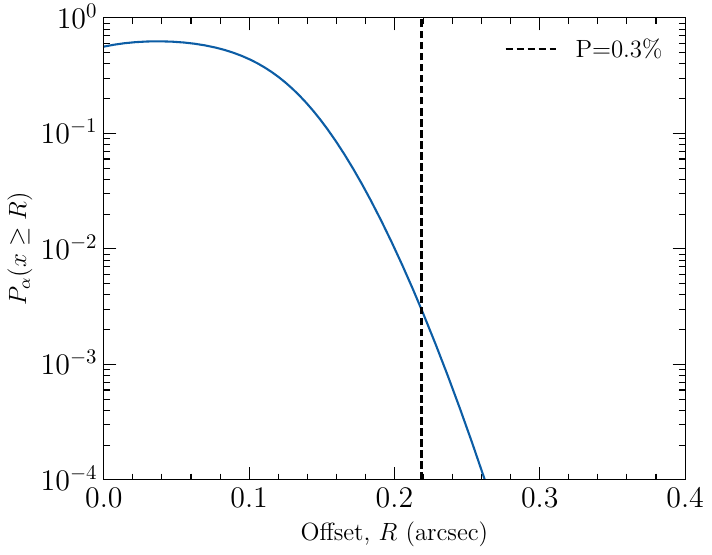}	
	\caption{Probability that a spatial offset greater than R is drawn from the Rayleigh component of the mixture distribution shown in Figure \ref{fig:HSCoffsets} instead of the exponential component. The offset where this probability is 0.3\% is shown with a dashed line.} 
	\label{fig:ratio}%
\end{figure}


\subsection{Background and motivation}

The HSC-SSP transient survey was a high-resolution, narrow-field, deep survey, with 0.17" pixel scales and single-epoch depths of $g\sim26.4$ in the 1.77 square degree ultra-deep fields, which were surveyed in $g$, $r$, $i$, $z$ and $y$ bands \citep{Aihara2022ThirdProgram}. At these depths, an estimated 60\% of sources were blended \citep{Bosch2018ThePipeline}. We now apply our scene-modeling approach to the characterization of variable AGN in deep multi-epoch imaging from the HSC-SSP transient survey. 

AGN studies are a key science driver for LSST, which is expected to observe on the order of 10 million AGN  \citep{DeCicco2021, Bianco2022}. A large fraction of Type 1 AGNs are variable: at least 90\% of Stripe-82 quasars of magnitude $g<20.5$ exhibit variability greater than $0.03$ mags \citep{Sesar2007, Barth2014}. Variability selection is therefore a key tool for quasar identification \citep[][e.g]{MacLeod2011,Butler2011} and has been found to produce major improvements in classification completeness compared to techniques based on color and morphology criteria alone in simulated LSST data \citep{Palanque2011,Savic2023,Burke2023DwarfSurveys}. 
Though variability enables highly complete AGN sample selection, optically variable AGN will also be a major source of contamination in searches for other transients: in the ZTF TDE alert filter, for example, mid-IR variability in the WISE survey was essential for removing AGN contaminants \citep{vanVelzen2021SeventeenStudies}. The presence of a bright and variable AGN can inhibit accurate extraction host galaxy SEDs, and therefore the estimation of photometric redshifts, which are important not only for cosmology studies but for identifying the hosts of transients such as GRBs and kilonovae \citep{Nugent2022}. 

As part of our HSC AGN analysis, we undertake a pilot study for the search for rare massive black holes (MBHs) that are spatially offset from their host galaxy nuclei, in preparation for a larger-scale analysis with LSST. Recent cosmological simulations that do not artificially tie MBHs to halo centers have found that the majority of MBHs of mass $M_{\text{BH}}<10^7M_{\odot}$ have a $>15$kpc spatial offset from their host nucleus, with the fraction increasing for decreasing mass \citep{Bellovary2019MultimessengerGalaxies,Ricarte2021UnveilingSignatures}. Variability-selection has proven to be particularly effective at identifying low mass AGN in dwarf galaxies where traditional spectroscopic diagnostics are less effective \citep[e.g.][]{Baldassare2018,Baldassare2020AFactory,Ward2022Variability-selectedWISE,Burke2022DwarfFields}, making it a promising approach for the identification of wandering MBHs. Forecasts for the number of $10^4 M_{\odot}<M_{\text{BH}}<10^6M_{\odot}$ MBHs with detectable variability in LSST range from 1,500 to 21,000 depending on seed mass and wandering fraction \citep{Burke2023DwarfSurveys}.

There are a variety of predictions for the accretion luminosities of such wandering AGN \citep{Pacucci2018Function,DiMatteo2023ANoon}, but even if they are not observable from typical gas accretion, they may be observable as tidal disruption events if the wandering MBHs retain their nuclear star clusters \citep[e.g][]{Lin2018ACluster}. The goal of this pilot study is to determine what spatial offsets will be detectable by applying multi-epoch scene modeling to LSST-like imaging data, using a sample of variable AGNs with well-determined spectroscopic redshifts and black hole masses \citep{Aihara2022ThirdProgram, Burke2024}.

\subsection{Light curves, host galaxy SEDs, and spatial offset distributions for variable AGN in HSC}

In this section we will analyze 542 variable AGN in the Hyper Suprime Cam imaging of the $\sim1.5$ deg$^2$ COSMOS-field \citep{Aihara2022ThirdProgram} without the need for reference images. A previous spectroscopic study of this sample found these AGNs to have redshifts up to $z\sim4$ and employed SED fitting to determine that the host galaxy stellar masses fell between $10^{8.5} M_{\odot}< M_*<10^{11}$; the BH mass distribution of this sample spans the range of $10^{7} M_{\odot}< M_*<10^{9.8}$ \citep{Burke2024}.  

We used the HSC \texttt{data-access-tools} to obtain 120$\times$120 arcsec cutouts of the multi-epoch warped (aligned and sky subtracted) HSC images and corresponding PSF images from the deep and ultra-deep fields in the Public HSC Data Release 3 \citep{Aihara2022ThirdProgram}. We obtained approximately 100 \textit{grizy} images for each AGN. We repeated the  \scarlettwo{} modeling procedure described for the ZTF TDEs in \autoref{sec:sne} with two major differences. Firstly, we did not apply any constraints on whether the transient is `on' or `off' because the variable AGN are always present to some extent. Secondly, because our primary motivation was the measurement of transient--host nucleus offsets and the host galaxies were very compact and faint, we instead enforced a S\'{e}rsic profile for the initial scene models, rather than applying a non-parametric galaxy model then fitting a S\'{e}rsic profile to the transient-subtracted image. 

In \autoref{fig:HSCeg}, we show an HSC observation, the \scarlettwo{} model for that observation derived from fitting the full multi-epoch data set, and the corresponding residual image for two COSMOS AGN: one in a compact, AGN-dominated galaxy -- which is very typical for this sample -- and one in an extended galaxy of comparable flux to the variable AGN. \scarlettwo{} produces high quality models and light curves for all AGN in the HSC sample. Three example light curves from this procedure are shown in \autoref{fig:HSClc}.

In \autoref{fig:SEDexamples}, we compare the host galaxy SED from the \scarlettwo{} model with the estimated host galaxy SED from a spectroscopic decomposition of the combined AGN and host spectrum obtained from \texttt{cigale} \citep{Boquien2019} fitting, as reported in \citet{Burke2024}. We show the results for a host galaxy-dominated system and an AGN-dominated system. \scarlettwo{} finds comparable galaxy SEDs to the \texttt{cigale} model results, even when only presented with $grizy$ images. In addition, the \scarlettwo{} sampling procedure enables the quantification of uncertainties in the galaxy SED arising from the presence of the AGN. 

We point out that in this comparison the \texttt{cigale} model may not be able to establish the ground truth for the host spectrum either.
However, the imaging-based decomposition we performed with \scarlettwo{} suffers from a different set of degeneracies. For extremely compact and high redshift galaxies, i.e. with insufficient extended emission from the galaxy to determine its SED from regions uncontaminated by the AGN, the AGN spectrum and host galaxy spectrum cannot be separated without additional assumptions---which \texttt{cigale} makes, but we have not. In such cases, \scarlettwo{} can produce light curves that accurately represent the variability, but cannot decide to which object it should attribute the minimum flux level of both sources. In such cases, including a single high-resolution space-based image with a more compact PSF would enable \scarlettwo{} to better distinguish between the host galaxy emission and the nuclear AGN emission because the AGN's effective area of influence is smaller. Future work will investigate the incorporation of existing \textit{Hubble Space Telescope} imaging of the COSMOS AGN \citep{Zhong2022AField} into \scarlettwo{} modeling for improved AGN-host decomposition. We will also investigate the option for spectral priors to aid this decomposition.

Our method can be applied to identify the presence of spatially offset `wandering' AGN in future LSST imaging. The distribution of host--AGN offsets (\autoref{fig:HSCoffsets}) peaks at around a quarter of a pixel, and is well-fit by a mixture distribution consisting of a Rayleigh distribution (expected for a population of offsets given some typical positional uncertainty) with $\sigma = 0.051$" and an exponential tail with exponent $\alpha=14.2$. The ratio between the two components of the mixture distribution was 1:0.56. In \autoref{fig:ratio} we show the probability that an offset greater than R is drawn from the Rayleigh component of the mixture distribution shown in Figure \ref{fig:HSCoffsets} instead of the exponential component. We find that our modeling strategy applied to LSST-like imaging will be sensitive to spatial offsets $>0.22$" when applying a 3$\sigma$ cutoff. Such objects should be investigated as potential candidates for spatially offset AGN. This distribution and offset cutoff is much smaller than that measured from the ZTF AGN sample, which was limited by the low resolution 1.0" pixel scale \citep{Ward2021AGNsFacility}.
As a result, in future work with larger AGN populations, differences in offset as a function of MBH mass could be compared to predictions of wandering fraction vs mass from cosmological simulations, where the majority of MBHs of mass $10^8M_\odot$ are expected to be non-nuclear.

\section{Technical Discussion}
We tested the run times when undertaking modeling of a series of 100 multi-epoch ZTF images of size 120 by 120 pixels, containing one TDE, one host galaxy, and two background galaxies. Using an AMD EPYC 7763 CPU on the NERSC cluster Perlmutter\footnote{\url{https://docs.nersc.gov/systems/perlmutter/architecture/}}, producing the full scene model using the setup described in Section 4 took an average of 329.6 s over 10 runs. By comparison, by switching to the 40GB Nvidia A100 GPU, this time was reduced to 52.6 s, producing a 6.3 times speed-up. For cutouts that are not pre-aligned to the same wcs, the rendering operation is repeated throughout the fitting, increasing the CPU processing time from 5.5 minutes to 3.5 hours. It is therefore recommended that for large multi-epoch datasets, the imaging be pre-aligned if possible to reduce computation time. We also note that the GPU time for MCMC sampling over all flux, position and morphology parameters took 4.5 hours on average. Future work will implement a Hessian vector product approximation of the Fisher matrix enabling fast approximation of parameter uncertainties, allowing users to avoid full MCMC sampling unless necessary for detailed analysis of the posteriors.

We re-emphasize that, in most situations, difference imaging provides a faster method of transient detection and an optimal light curve for characterizing variability relative to a reference. However, a full scene model is advantageous when joint modeling of the host galaxy is required -- such as for the measurement of transient-host offsets -- and when the transient or AGN is already present in the reference imaging. A scene modeling approach provides a convenient way to jointly model imaging data taken with different telescopes at different times, enabling the production of light curves across multiple surveys without concern for reference image mismatch. The ability to simultaneously sample the posteriors of both transient and host galaxy parameters can be beneficial for detailed characterization of sources of interest.

In the HSC AGN analysis presented in Section 5, we chose to enforce a S\'{e}rsic profile for the galaxy morphologies when fitting. Adding this stronger constraint on galaxy morphology is only recommended for compact AGN host galaxies where there are strong degeneracies between flux from the variable AGN and the host. This constraint is not required for nuclear transients where pre-flare imaging is available, or for host galaxies where the extended galaxy emission beyond the central AGN is well-resolved in at least one image, enabling unambiguous measurement of the host galaxy SED. A carefully constructed prior for high redshift AGN host galaxies could also be used to avoid the need for a S\'{e}rsic profile constraint in this test case. We also note that for the AGN variability science cases described in Section 5.1, systematic flux errors introduced in the AGN light curve via the enforcement of a S\'{e}rsic profile do not inhibit the detection of the AGN variability itself, or analysis of the variability timescale.

\section{Summary and conclusions}
We have presented a time-domain extension to the \scarlettwo{} scene modeling and deblending code. It can model multi-epoch, multi-band, multi-resolution imaging data without the need for reference or difference images and constrain source fluxes as either time-varying or `static' to assist in model fitting. We demonstrate how the method can be used to produce models of variable point sources and their host galaxies, enabling extraction of transient light curves and positions, and host galaxy morphology models and SEDs. The method has been applied to simulated supernova imaging at LSST-like resolutions, low-resolution ZTF imaging of tidal disruption events, and variable AGNs in the HSC-SSP transient survey. The automated initialization procedures and the option to use prior-informed non-parametric galaxy models provide a high level of flexibility  for various time-domain studies. The ability to carry out statistical sampling over source parameters in image space means that \scarlettwo{} can be used for careful quantification of uncertainties in source fluxes, positions, and host morphologies when studying systems of particular interest. We demonstrate that \scarlettwo{} can be used to confirm the spatial offsets of transients and their host galaxies for classification of nuclear and non-nuclear transients, and to identify populations of AGN `wandering' away from their host galaxy nuclei. 

This method will be most powerful when 
high cadence time-domain ground-based imaging from surveys like ZTF and LSST is jointly modeled with higher resolution space-based imaging from the \textit{Hubble Space Telescope}, \textit{Euclid} \citep{Racca+16}, and the \textit{Roman Space Telescope} \citep{Akeson+19}.
The option for GPU acceleration means that image processing with \scarlettwo{} can be run at scale. 
We have made the \scarlettwo{} code available for the wider
astrophysical community at \url{https://github.com/pmelchior/scarlet2}. Example jupyter notebooks for the methods described in each section are available at \url{https://github.com/charlotteaward/TimeDomainScarlet2}.

\section{Acknowledgements}
We thank the anonymous reviewers for their helpful feedback that was very beneficial to the paper. Charlotte Ward acknowledges support from the LSST Discovery Alliance under grant AWD1008640 (PI: Ward).  

\bibliographystyle{elsarticle-harv} 
\bibliography{references,referencesP2,main}






\end{document}